\documentclass{article}
\usepackage{microtype,graphicx,booktabs}
\usepackage{amsmath,amssymb,array,multirow}
\usepackage{xcolor,enumitem,xurl,hyperref}
\usepackage[accepted]{mlsys2025}
\makeatletter
\renewcommand{\Notice@String}{}
\renewcommand{\mlsys@appearing}{}
\makeatother
\hypersetup{pdfsubject={LLM serving and cluster efficiency}}
\DeclareRobustCommand{\idx}[1]{(\textbf{\underline{#1}})}
\definecolor{findingblue}{HTML}{08519C}
\definecolor{findingbg}{HTML}{EDF3FA}
\usepackage[framemethod=default]{mdframed}
\newcommand{\finding}[2]{%
  \begin{mdframed}[linecolor=findingblue,backgroundcolor=findingbg,linewidth=0.5pt,roundcorner=0pt,leftmargin=0pt,rightmargin=0pt,innerleftmargin=5pt,innerrightmargin=5pt,innertopmargin=5pt,innerbottommargin=5pt,skipabove=0.5em,skipbelow=0.5em,splittopskip=5pt,splitbottomskip=5pt,nobreak=true]
  \noindent\textcolor{findingblue}{\textbf{Finding #1.}} #2
  \end{mdframed}}
\newcommand{\fkey}[1]{\textcolor{findingblue}{\textbf{#1}}}
\newcommand{\figinclude}[2]{\IfFileExists{figures/#1.pdf}{\includegraphics[width=#2]{figures/#1.pdf}}{\fbox{\parbox{.9\linewidth}{Figure in preparation: \texttt{\detokenize{#1}}.}}}}
\mlsystitlerunning{Introducing Cluster Efficiency to LLM Serving}
\begin{document}
\twocolumn[
\mlsystitle{Introducing Cluster Efficiency to LLM Serving: Pooling Economical Accelerators over Fast Interconnects}
\begin{mlsysauthorlist}
\mlsysauthor{Youhe Jiang}{cam}
\mlsysauthor{Pavle Glu\v{s}ac}{tt}
\mlsysauthor{Binhang Yuan}{hkust}
\mlsysauthor{Eiko Yoneki}{cam}
\end{mlsysauthorlist}
\begin{center}
\small
\textsuperscript{1}University of Cambridge\qquad
\textsuperscript{2}Tenstorrent\qquad
\textsuperscript{3}HKUST
\end{center}
\mlsyskeywords{LLM inference, accelerator interconnects, energy efficiency, serving cost}
\vskip 0.3in
\begin{abstract}
Long-context large language model (LLM) serving places substantial demands on compute, memory capacity, and memory bandwidth.
Economical accelerators can offer these resources at lower cost, but their limited per-device capacity requires pooling across devices, making communication overhead critical.
We study cluster efficiency, measured in tokens per dollar and per joule, and show how low-latency, high-bandwidth interconnects enable economical accelerators to retain their cost advantage.
Guided by a characterization of coding-agent workloads and accelerator scaling, we realize this design on the Tenstorrent Galaxy, a cluster of Blackhole chips.
The serving stack combines integrated communication engines, pipelined ring transfers, communication-computation overlap, and device-side decode replay to reduce collective overhead.
Across three dense models in BF16 and 8-bit configurations with 8K-256K-token contexts, the Galaxy achieves 1.07-2.24$\times$ the tokens per dollar of an eight-GPU RTX PRO 6000 Blackwell server under a three-year acquisition-and-electricity cost model.
Emulating TCP-class communication on the same chips reduces throughput by 23.5-67.4\% and eliminates the cost advantage in every evaluated configuration.
These results demonstrate how fast interconnects translate the per-dollar resource advantages of economical accelerators into cost-efficient long-context LLM serving.
\end{abstract}
]

\section{Introduction}
\label{sec:intro}

Large language models (LLMs) now power a widening range of applications, from conversational assistants to coding agents, and the models behind them have grown to tens and hundreds of billions of parameters~\cite{qin2025mooncake,zhu2026tracelab,grattafiori2024llama3,yang2025qwen3}.
A model of this size outgrows a single device, so efficient LLM serving typically runs it over a cluster of accelerators, whether GPUs, TPUs, or NPUs~\cite{pope2023efficiently,jouppi2023tpuv4,vasiljevic2021tenstorrent}.
These accelerators are not interchangeable: they differ in compute throughput, memory bandwidth, memory capacity, and power draw~\cite{choquette2023h100,lenovo2026b200,nvidia2025rtxpro6000}.
Which accelerator a serving cluster is built from is therefore a first-order decision for its efficiency.

Powerful GPUs such as the B200, H200, and H100 provide high per-device memory capacity and bandwidth (device efficiency), so a large model spans few of them and the links between them are short and fast~\cite{nvidia2023h100,nvidia2024h200,lenovo2026b200}.
Low-end GPUs such as the RTX A5000 and RTX 4090 provide much less of both~\cite{nvidia2021rtxa5000,nvidia2022rtx40}.
A large model and its KV cache must therefore span several machines of them, and those machines typically reach one another over poor connections, which bottleneck the serving process~\cite{shoeybi2019megatron,narayanan2021megatron2}.
Our characterization confirms both halves of this picture: a machine of such accelerators lacks the memory and compute for a large model and its KV cache, and the links between such machines are far slower than NVLink (\S\ref{sec:characterization}, \textbf{Finding~2}).
Under this view, a serving cluster is built from the fewest, highest-end GPUs that the model allows.

However, our characterization (\S\ref{sec:characterization}) shows that this preference does not always hold once the cluster is judged by what it costs.
In tokens per dollar and tokens per joule, the high-end part is not necessarily the better choice, because its higher per-device capacity comes with a higher price and a higher power draw~\cite{cnbc2024b200,jarvislabs2025h200,lenovo2026b200,nvidia2023h100}.
Normalized by price, the ranking can reverse (\S\ref{sec:characterization}, \textbf{Finding~3}): per \$1000, a consumer RTX 4090~\cite{nvidia2022rtx40} supplies 413 peak 8-bit TFLOPS of compute, more than three times the 129 TFLOPS of a B200~\cite{cnbc2024b200}. The weakness of such a lower-end accelerator lies in its per-device capacity rather than in what it offers for each dollar spent.

This observation motivates us to \textbf{rethink} the design of the serving cluster.
Rather than asking which device is the fastest (\textit{device efficiency}), we ask how to build a cluster that serves the most tokens per dollar and per joule~\cite{griggs2024melange,stojkovic2025dynamollm}, a property that we define as the \emph{cluster efficiency} of the deployment.

To reach cluster efficiency, we propose a counter-intuitive design: build the cluster from lower-end accelerators, which offer more capacity per dollar and per watt, and connect them with fast links instead of the slow cross-machine connections they are typically given, so that the communication bottleneck of pooling them disappears.
Our characterization demonstrates that over many fast links per chip, small memories and compute pool at little cost, with capacity, prefill, and decode scaling with the chip count while collectives stay a small share of each step, whereas over weak links the same chips lose most of the speedup (\S\ref{sec:characterization}, \textbf{Finding~4}).
Once the collectives (i.e., model-parallel communication~\cite{shoeybi2019megatron,narayanan2021megatron2}) are cheap, the small memories and compute of many lower-end chips accumulate into a pool\footnote{A pool is the set of devices that serves one model replica over tensor parallelism; a cluster holds one or more pools and is the unit that a provisioner has to buy and power.} that keeps both advantages, the per-dollar and per-watt capacity of each chip and the scalability that today belongs to tightly linked high-end clusters.
Our characterization further shows that, pooled this way, lower-end chips overtake high-end GPUs of the same tensor-parallel degree from as few as two devices in both tokens per dollar and tokens per joule (\S\ref{sec:characterization}, \textbf{Finding~5}).

We realize this design as a communication system over the chips of the Tenstorrent Blackhole Galaxy~\cite{tenstorrent2026galaxy,tenstorrent2026blackhole}, built for cluster efficiency (\S\ref{sec:design}).
Each chip is a lower-end part on its own, with a fraction of the compute and memory of a high-end GPU, and the communication system is what turns tens to hundreds of them into one efficient pool: it connects the chips over links integrated on each chip (\S\ref{sec:design:galaxy}), carries a transfer across the ring with little overhead on each hop (\S\ref{sec:design:comm}), and shapes model inference around those links so that the links alone set the cost of every collective (\S\ref{sec:design:pool}).
\S\ref{sec:implementation} describes its implementation in the serving stack, from the model code down to the links and the Ethernet-core firmware beneath them.

To our knowledge, this paper is the \textit{first} to introduce the concept of cluster efficiency and to rethink the design of the LLM serving cluster on top of it, pooling lower-end accelerators over fast links in place of fewer high-end GPUs.
Our contributions are summarized as follows:

\noindent\textbf{\underline{Contribution 1:}} We characterize long-context agent serving on lower-end and high-end accelerators (\S\ref{sec:characterization}).
Five findings (\textbf{Findings~1-5}) show that \idx{i} long-context agent serving needs compute, memory capacity, and memory bandwidth together; \idx{ii} a machine of lower-end accelerators lacks them, and the links between such machines are slow; \idx{iii} per dollar, lower-end accelerators can offer more memory, bandwidth, and compute than high-end GPUs; \idx{iv} over fast links, lower-end chips pool at little cost, whereas over weak links they lose most of the speedup; and \idx{v} pooled over fast links, lower-end chips beat high-end GPUs in tokens per dollar and per joule.
These characterizations motivate the cluster design of utilizing fast links to connect lower-end accelerators for achieving cluster efficiency.

\noindent\textbf{\underline{Contribution 2:}} Motivated by the characterization results, we define cluster efficiency and realize a communication system on the Tenstorrent Blackhole Galaxy~\cite{tenstorrent2026galaxy}, a cluster of lower-end accelerators (\S\ref{sec:design}, \S\ref{sec:implementation}).
The communication system links the chips through ports integrated on every chip (\S\ref{sec:design:galaxy}), carries transfers across the ring with little overhead on each hop (\S\ref{sec:design:comm}), and shapes model inference around those links (\S\ref{sec:design:pool}), implemented throughout the Galaxy serving stack, from the vLLM serving engine and runtime down to the collective kernels (\S\ref{sec:implementation}).

\noindent\textbf{\underline{Contribution 3:}} We evaluate the Galaxy against an RTX PRO 6000 Blackwell server on three dense models, two precisions, and 8K-256K contexts (\S\ref{sec:evaluation}).
The Galaxy delivers 1.07-2.24$\times$ the tokens per dollar and 1.00-2.27$\times$ the tokens per joule of the GPU server across all configurations (\S\ref{sec:rq1}).
An ablation of the Galaxy chips connected over slow links cuts their throughput by 23.5-67.4\% and erases every one of its cost-efficiency wins over the GPU server (\S\ref{sec:rq2}).

\section{Background}
\label{sec:background}

\vspace{0.2em}\noindent \textbf{LLM serving.}
Serving a decoder-only LLM runs each request through two phases: prefill, which processes the prompt in one forward pass and writes its KV cache, and decode, which generates one token per step while reading the KV cache of every earlier token~\cite{pope2023efficiently,kwon2023pagedattention}.
Serving engines batch many requests into one decode step so that each weight read serves all of them, admit and retire requests between steps, and page the KV cache so that memory is allocated per block rather than per maximum context~\cite{yu2022orca,kwon2023pagedattention,zheng2024sglang,sun2024llumnix}; chunked prefill schedules prompt chunks alongside decode, and prefill and decode may be split across machines~\cite{agrawal2024sarathi,zhong2024distserve,patel2024splitwise,qin2025mooncake}.
The cost of serving is therefore set by how much memory and bandwidth a deployment holds and by how many requests share them at each step, and agentic workloads raise both with long sessions whose cached state stays resident~\cite{zhu2026tracelab}; this paper judges a serving cluster by what that capacity costs per token served, in dollars and in joules, which we call its cluster efficiency, rather than by what any one of its devices delivers on its own.

\vspace{0.2em}\noindent \textbf{Parallelism.}
A model that exceeds one accelerator is split by parallelism.
Tensor parallelism shards every layer across devices and synchronizes them at each layer with collectives, so its step time depends on the latency and bandwidth of the links between the devices~\cite{shoeybi2019megatron,narayanan2021megatron2}; pipeline parallelism assigns contiguous layers to each device and passes only activations between stages, which tolerates slower links at the cost of pipeline bubbles~\cite{huang2019gpipe,narayanan2021megatron2}; expert and sequence parallelism partition the mixture-of-experts layers and the token dimension~\cite{lepikhin2021gshard,korthikanti2023sequence,liu2024ringattention}.
Inference systems combine these with data parallelism across model replicas and choose the combination per model and hardware~\cite{aminabadi2022deepspeed,zheng2022alpa,pope2023efficiently}.
For serving, tensor parallelism is the way to pool the memory and bandwidth of several devices behind one model replica, and the collectives it issues at every layer are the price of pooling; this paper shows that this price, set by the links rather than by the devices, decides whether the per-dollar advantage of lower-end accelerators survives pooling, and builds a communication system that keeps it small.

\vspace{0.2em}\noindent \textbf{Heterogeneous resources.}
The hardware that serves LLMs is no longer one kind of device.
Within one vendor's line, successive generations differ in memory, bandwidth, and interconnect, from H100 and H200 to B200~\cite{choquette2023h100,nvidia2024h200,lenovo2026b200}, and the same generation ships in data-center and workstation parts that differ mostly in links and price~\cite{nvidia2025rtxpro6000}.
Beyond GPUs, TPUs~\cite{jouppi2023tpuv4}, wafer-scale and dataflow accelerators~\cite{lie2023cerebras,abts2022groq}, and Tenstorrent's Ethernet-connected chips~\cite{vasiljevic2021tenstorrent,tenstorrent2026blackhole} trade per-device capability for cost and scale-out.
Serving systems have begun to exploit this heterogeneity: by mixing GPU types across replicas~\cite{griggs2024melange,jiang2025thunderserve}, by placing one model across devices of unequal memory and links~\cite{jiang2023hexgen,jiang2025hexgen,mei2025helix,zhang2025cauchy,wei2026hmaserve}, by using preemptible or consumer devices~\cite{miao2024spotserve,borzunov2023petals,sheng2023flexgen}, and by characterizing new accelerators for decode~\cite{lim2027chippipe,rhee2025hpu}.
These systems take the devices as given and schedule serving around their differences; this paper rethinks the cluster design itself, and shows that lower-end accelerators connected over fast links beat high-end GPUs on the two metrics a provisioner ultimately pays for, tokens per dollar and tokens per joule.
\nocite{jiang2023hexgen,jiang2025demystifying,jiang2025thunderserve,jiang2025hexgen,yanhexiscale,tong2025parallax,peng2025hexgen,jiang2026boute,yan2025areal,yao2026opentela,jianghexgen}

\section{Characterization}
\label{sec:characterization}
This section builds the case for cluster efficiency in five steps, each closed by a finding.
We first characterize what long-context agent serving demands from hardware, then show why lower-end accelerators look unfit for it one device at a time, why they nevertheless offer more resources per dollar, how many fast links per chip pool their small memories and compute at little cost, and that the resulting pool beats high-end GPUs per dollar and per joule.
The workload is the TraceLab coding-agent trace~\cite{zhu2026tracelab}; the scaling experiments run one Llama-3.1-8B replica at 32K context on one, two, four, or eight devices of the Galaxy and of the GPU server (\S\ref{sec:evaluation} describes both machines).

\vspace{0.2em}\noindent \textbf{Agent serving needs compute and memory.}
We analyze 357149 model invocations from 4265 TraceLab coding-agent sessions (Table~\ref{tab:char:workload}); an invocation is one agent step, not a complete user request.
The median step adds 875 fresh input tokens to a cached prefix of 118656 tokens and generates 214 output tokens, so cached prefixes supply 95.7\% of the context, and the 99th-percentile step holds 822889 tokens.
The two phases load different resources.
Prefill is compute-bound: one 32K prompt sustains 489 TFLOP/s in FP8 on one GPU, 49\% of its dense peak, and 45 TFLOP/s on one Galaxy chip.
Decode is memory-bound: the GPU's effective decode bandwidth, modeled weight and KV bytes divided by measured step time, rises from 1.22 TB/s at one user to 1.48 TB/s at 37 users, 93\% of its nominal 1.6 TB/s.
The resident state is large: mapping each session's peak context to Llama-3.3-70B BF16 KV gives a median of 22.6 GB and a 90th percentile of 73.9 GB before weights, and the 36 sessions open at the busiest moment hold 9.6 million context tokens, 3.15 TB of KV cache under that model.

\begin{table}[ht]
\centering
\caption{TraceLab agent-step demand: tokens per step over 357149 steps, and the Llama-3.3-70B BF16 KV cache at each session's peak context (decimal GB, before weights and runtime buffers).}
\label{tab:char:workload}
\setlength{\tabcolsep}{3pt}
\renewcommand{\arraystretch}{1.0}
\resizebox{\columnwidth}{!}{%
\small\begin{tabular*}{\columnwidth}{@{\extracolsep{\fill}}l | r | r | r@{}}
\hline
 & P50 & P90 & P99 \\
\hline
Fresh input (tokens) & 875 & 7119 & 144688 \\
\hline
Cached prefix (tokens) & 118656 & 247836 & 816627 \\
\hline
Output (tokens) & 214 & 1169 & 5050 \\
\hline
Context held in KV (tokens) & 124024 & 256770 & 822889 \\
\hline
Session peak KV, 70B BF16 (GB) & 22.6 & 73.9 & 139.3 \\
\hline
\end{tabular*}}
\end{table}

\finding{1}{Long-context agent serving needs \fkey{compute} for prefill and \fkey{memory} capacity and bandwidth for decode and for a resident KV cache that grows with every open session. A serving pool must supply both.}

\vspace{0.2em}\noindent \textbf{Lower-end machines lack memory, compute, and links.}
Conventional provisioning keeps large-model serving on high-end GPUs because a machine of lower-end accelerators falls short in two ways (Table~\ref{tab:char:devices}).
Within the machine, memory and compute are small: an eight-GPU RTX A5000 server holds $8\times24$ GB and an RTX 5090 card 32 GB, against 80-180 GB per H100 or B200, and their peak compute is a fraction of the high-end parts'.
The retained Llama-3.3-70B dimensions imply 141.1 GB of raw BF16 weights before the 73.9 GB session KV of \textbf{Finding 1}, so a long-context pool of such parts must span machines.
Across machines, the links are slow: the A5000 server reaches other servers at 25 GB/s of RDMA, about 3 GB/s per GPU, while high-end GPUs share 450-900 GB/s of NVLink inside one chassis.
Pooling many small machines over slow links is exactly what conventional provisioning avoids, so lower-end parts are passed over for large models at any price.

\begin{table}[ht]
\centering
\caption{Per-device resources of high-end and lower-end accelerators~\cite{lenovo2026b200,nvidia2024h200,nvidia2023h100,nvidia2023l40s,nvidia2025rtxpro6000,nvidia2021rtxa5000,nvidia2025rtxblackwell}. 8-bit rates are dense vendor peaks (GeForce FP8 with FP16 accumulation; the RTX PRO 6000 Blackwell rate is derived from its 4 PFLOPS sparse FP4 rating). Link: the fastest link to another device in the same machine, per direction (NVL: NVLink generation; PCIe: PCIe generation). $^\dagger$FP16 is listed, because the RTX A5000 has no FP8 support in hardware.}
\label{tab:char:devices}
\setlength{\tabcolsep}{3pt}
\renewcommand{\arraystretch}{1.0}
\resizebox{\columnwidth}{!}{%
\small\begin{tabular*}{\columnwidth}{@{\extracolsep{\fill}}l | r | r | r | r | l@{}}
\hline
Accelerator & Price \$ & GB & GB/s & TFLOPS & Link, GB/s \\
\hline
B200 & $\sim$35000 & 180 & 7700 & 4500 & NVL5, 900 \\
\hline
H200 & $\sim$35000 & 141 & 4800 & 1979 & NVL4, 450 \\
\hline
H100 SXM & $\sim$27500 & 80 & 3350 & 1979 & NVL4, 450 \\
\hline
L40S & $\sim$7500 & 48 & 864 & 733 & PCIe4, 32 \\
\hline
RTX PRO 6000 & 8565 & 96 & 1597 & $\sim$1000 & PCIe5, 63 \\
\hline
RTX A5000 & 2250 & 24 & 768 & 111$^\dagger$ & PCIe4, 32 \\
\hline
RTX 4090 & 1599 & 24 & 1008 & 661 & PCIe4, 32 \\
\hline
RTX 5090 & 1999 & 32 & 1792 & 838 & PCIe5, 63 \\
\hline
\end{tabular*}}
\end{table}

\finding{2}{A machine of lower-end accelerators has too little \fkey{memory} and \fkey{compute} for a large model and its KV cache, and the \fkey{links} between such machines are slow. Serving on them means pooling several such machines, and slow links between the machines are what prevent it.}

\vspace{0.2em}\noindent \textbf{Per dollar, lower-end could offer more.}
Normalized by price, the ranking reverses (Table~\ref{tab:char:perdollar}).
Per \$1000, a Blackhole p150a supplies 22.9 GB of memory, 366 GB/s of memory bandwidth, and 475 peak Block FP8 TFLOPS, against 5.1 GB, 220 GB/s, and 129 TFLOPS for a B200 and 11.2 GB, 186 GB/s, and 117 TFLOPS for an RTX PRO 6000 Blackwell.
The consumer RTX 4090 and RTX 5090 follow the same pattern with 15.0 and 16.0 GB, 630 and 896 GB/s, and 413 and 419 TFLOPS, about three to four times the B200 on every axis, while the data-center H100 and H200 sit lowest at 2.9 and 4.0 GB, 122 and 137 GB/s, and 72 and 57 TFLOPS.
Prices are the midpoint of NVIDIA's \$30-40K estimate for the B200~\cite{cnbc2024b200}, midpoints of reported ranges for the H200 and H100~\cite{jarvislabs2025h200}, a reported L40S card price~\cite{modal2025l40s}, the RTX PRO 6000 Blackwell price~\cite{punch2026rtxpro6000server,bytehaus2026aiservers}, launch prices for the A5000, RTX 4090, and RTX 5090~\cite{cgchannel2021ampere,nvidia2022rtx40,nvidia2025rtx50}, and the p150a list price~\cite{tenstorrent2026blackhole}.
The 8-bit peaks cover different number formats and accumulation modes, so they are an upper bound on, rather than a prediction of, the compute that each device delivers on a real workload.

\begin{table}[ht]
\centering
\caption{Memory (GB), memory bandwidth (GB/s), and peak 8-bit compute (TFLOPS) per \$1000, at the prices of Table~\ref{tab:char:devices} and the p150a's \$1399 list price~\cite{tenstorrent2026blackhole}.}
\label{tab:char:perdollar}
\setlength{\tabcolsep}{3pt}
\renewcommand{\arraystretch}{1.0}
\resizebox{\columnwidth}{!}{%
\small\begin{tabular*}{\columnwidth}{@{\extracolsep{\fill}}l | r | r | r@{}}
\hline
Accelerator & GB & GB/s & TFLOPS \\
\hline
B200 & 5.1 & 220 & 129 \\
\hline
H200 & 4.0 & 137 & 57 \\
\hline
H100 SXM & 2.9 & 122 & 72 \\
\hline
L40S & 6.4 & 115 & 98 \\
\hline
RTX PRO 6000 & 11.2 & 186 & 117 \\
\hline
RTX A5000 & 10.7 & 341 & 49$^\dagger$ \\
\hline
RTX 4090 & 15.0 & 630 & 413 \\
\hline
RTX 5090 & 16.0 & 896 & 419 \\
\hline
Blackhole p150a & 22.9 & 366 & 475 \\
\hline
\end{tabular*}}
\end{table}

\finding{3}{Per dollar, lower-end accelerators could offer more \fkey{memory}, \fkey{memory bandwidth}, and \fkey{compute} than high-end GPUs. Their weakness is per device, not per dollar; the comparison that matters is per dollar.}

\vspace{0.2em}\noindent \textbf{Link speed decides whether pooling pays.}
We vary tensor parallelism at a fixed model and context and report each quantity relative to one chip, over the Galaxy's own links and over the same chips with emulated TCP-class links (Table~\ref{tab:char:pooling}).
Over the native links, the 8-bit active batch grows from 10 users on one chip to 25, 54, and 113 users on two, four, and eight: capacity grows faster than the chip count because the fixed weight footprint spreads over more memory.
The pooled compute speeds up a fixed workload as well: a 32K prefill falls from 18.04 s to 3.11 s (5.80$\times$) and the eight-user decode step from 99.1 ms to 16.8 ms (5.90$\times$), while collectives take only 4.8-6.5\% of a profiled decode step.
Over TCP-class links the same chips keep the same capacity but lose most of the speedup: the 32K prefill on eight chips takes 8.20 s (2.20$\times$), the eight-user step 42.0 ms (2.36$\times$), and the gain stalls beyond four chips because every collective now pays a link latency of 100-250 $\mu$s and communication fills 63\% of the step.
The capacity is still there, but the added chips mostly wait on one another; \S\ref{sec:design:objective} names the terms behind this loss and \S\ref{sec:rq2} measures them.

\begin{table}[ht]
\centering
\caption{Llama-3.1-8B 8-bit at 32K on two, four, and eight Galaxy chips, relative to one chip, over fast links (the communication system of \S\ref{sec:design}) and over the same chips with emulated TCP-class links (the bandwidth and 64 KB latency of NCCL over TCP sockets, \S\ref{sec:setup}). Active users are the memory maximum; speedups are for a fixed workload (one 32K prefill, a one- or eight-user decode step). Collective share is the fraction of a decode step spent in collectives, measured with the device profiler over the fast links and computed from the step-time increase over TCP-class links.}
\label{tab:char:pooling}
\setlength{\tabcolsep}{3pt}
\renewcommand{\arraystretch}{1.0}
\resizebox{\columnwidth}{!}{%
\small\begin{tabular*}{\columnwidth}{@{\extracolsep{\fill}}l | r | r | r@{}}
\hline
Metric & TP2 & TP4 & TP8 \\
\hline
Active users (any link) & 2.50$\times$ & 5.40$\times$ & 11.30$\times$ \\
\hline
Native: 32K prefill speedup & 1.87$\times$ & 3.37$\times$ & 5.80$\times$ \\
\hline
Native: 1-user step speedup & 2.72$\times$ & 4.54$\times$ & 6.95$\times$ \\
\hline
Native: 8-user step speedup & 2.13$\times$ & 3.69$\times$ & 5.90$\times$ \\
\hline
Native: collective share & 5.2\% & 4.8\% & 6.5\% \\
\hline
TCP-class: 32K prefill speedup & 1.50$\times$ & 2.05$\times$ & 2.20$\times$ \\
\hline
TCP-class: 1-user step speedup & 2.31$\times$ & 2.61$\times$ & 2.50$\times$ \\
\hline
TCP-class: 8-user step speedup & 1.87$\times$ & 2.31$\times$ & 2.36$\times$ \\
\hline
TCP-class: collective share & 16.8\% & 40.4\% & 62.6\% \\
\hline
\end{tabular*}}
\end{table}

\finding{4}{Over many \fkey{fast links} per chip, small memories and compute \fkey{pool} at little cost: capacity, prefill, and decode scale with the chip count while collectives stay a small share of each step. Over \fkey{weak links}, the same chips keep their capacity but lose most of the speedup.}

\vspace{0.2em}\noindent \textbf{Pool size reverses the ranking.}
We compare each Galaxy replica with the GPU replica of the same tensor-parallel degree at its maximum batch, at \$5000 per chip and \$28137 per GPU over three years at \$0.15/kWh, with the power of the replica's own devices (Table~\ref{tab:char:pool-efficiency}).
One chip loses to one GPU: in 8-bit it delivers 0.63$\times$ the tokens per joule and 0.52$\times$ the tokens per dollar.
The pool reverses the ranking at two devices and widens it from there, reaching 1.25$\times$ in tokens per joule and 1.54$\times$ in tokens per dollar at eight.
The Galaxy replica gains because its users per chip keep rising, from 10 to 14 across the sweep, while the GPU replica's tokens per dollar peak at two GPUs and fall at eight, where collectives take half of its decode step.
BF16 shows the same reversal, reaching 1.32$\times$ and 1.59$\times$ at eight devices.
Several smaller replicas are another valid use of eight GPUs, and the main evaluation adopts them; the device ranking and the pool ranking disagree, and the cluster, not the device, is the unit a provisioner has to buy and power.

\begin{table}[ht]
\centering
\caption{Efficiency of one Llama-3.1-8B replica at its maximum batch (32K) on $n$ Galaxy chips over fast links (the communication system of \S\ref{sec:design}) against $n$ RTX PRO 6000 Blackwell GPUs on their own links: tokens per dollar in thousands (prices are given in \S\ref{sec:setup}) and the Galaxy/GPU ratios in tokens/\$ and tokens/J. The power of each device is profiled with its vendor's management interface during every timed round~\cite{tenstorrent2026ttsmi,nvidia2025nvml}.}
\label{tab:char:pool-efficiency}
\setlength{\tabcolsep}{3pt}
\renewcommand{\arraystretch}{1.0}
\resizebox{\columnwidth}{!}{%
\small\begin{tabular*}{\columnwidth}{@{\extracolsep{\fill}}l | r | r | r | r | r@{}}
\hline
Precision & $n$ & Galaxy tok/\$ & GPU tok/\$ & tok/\$ ratio & tok/J ratio \\
\hline
8-bit & 1 & 1020K & 1975K & 0.52$\times$ & 0.63$\times$ \\
\hline
8-bit & 2 & 2090K & 2011K & 1.04$\times$ & 1.07$\times$ \\
\hline
8-bit & 4 & 2317K & 1926K & 1.20$\times$ & 1.16$\times$ \\
\hline
8-bit & 8 & 2367K & 1533K & 1.54$\times$ & 1.25$\times$ \\
\hline
BF16 & 1 & 817K & 904K & 0.90$\times$ & 0.97$\times$ \\
\hline
BF16 & 2 & 1006K & 983K & 1.02$\times$ & 1.00$\times$ \\
\hline
BF16 & 4 & 1166K & 979K & 1.19$\times$ & 1.14$\times$ \\
\hline
BF16 & 8 & 1257K & 792K & 1.59$\times$ & 1.32$\times$ \\
\hline
\end{tabular*}}
\end{table}

\finding{5}{Pooled over fast links, lower-end chips beat high-end GPUs in \fkey{tokens per dollar} and reported \fkey{tokens per joule}, and the edge \fkey{grows with the pool}. Efficiency must be compared across whole pools, not single devices.}

\section{Cluster Efficiency and Its Realization on Galaxy}
\label{sec:design}
This section turns the five findings into a definition and shows how Galaxy realizes it.
\S\ref{sec:design:objective} defines cluster efficiency by its two metrics and the terms through which communication lowers it.
\S\ref{sec:design:galaxy} presents Galaxy, a pooled lower-end system, and the resources and links of its pools.
\S\ref{sec:design:comm} demonstrates its communication system, the links and routers that keep the communication overhead small.
\S\ref{sec:design:pool} shows how Galaxy adapts model inference to its fast links to improve cluster efficiency on real inference workloads.

\subsection{Definition and Communication Impact}
\label{sec:design:objective}
This subsection defines cluster efficiency and the terms through which communication lowers it.

\vspace{0.2em}\noindent \textbf{Objectives for cluster efficiency.}
Cluster efficiency has two objectives, tokens per joule and tokens per dollar, and both take the pool rather than the device as their unit (\textbf{Finding 5}).
For a pool that decodes $R$ tokens/s for active requests while its devices draw $P$ watts, over an active lifetime of $H$ seconds, at an acquisition cost of $C$ dollars and an electricity price of $e$ dollars per kWh,
\begin{equation}
\scalebox{0.92}{$\displaystyle
E=\frac{R}{P},\qquad
D=\frac{RH}{C+kP},\qquad k=\frac{eH}{3.6\times10^{6}}.$}
\label{eq:cost}
\end{equation}
Here $E$ is in tokens/J and $D$ in tokens/\$; $R$ counts tokens of active requests only, so padded slots do not inflate it; $P$ is the measured device power, whose telemetry boundary (chip, board, or system) must be stated with every value of $E$; and $k$ converts watts into lifetime electricity dollars, so the denominator of $D$ is the purchase price plus the electricity the pool draws over its life, and the numerator is every token it produces in that time.
The paper uses $H=3$ years of continuous use and $e=\$0.15$/kWh, under which electricity is under a tenth of the denominator, so $D$ is set mainly by throughput per dollar of hardware. This denominator is the three-year total cost of ownership that accelerator designers optimize, capital amortized over three years plus the electricity of operation~\cite{jouppi2021tpuv4i,barroso2018datacenter}, and the cost per generated token that serving-hardware studies report~\cite{peng2023chipletcloud}.
Both are provisioning metrics for comparing hardware, not predictions of utilization.

\vspace{0.2em}\noindent \textbf{Communication impact on cluster efficiency.}
Communication lowers both objectives through the time a decode step spends in collectives, and that time has two parts.
Following the standard $\alpha$-$\beta$ model, the collectives of one step cost
\begin{equation}
\scalebox{0.92}{$\displaystyle
T_{\mathrm{comm}}\approx\sum_{i=1}^{K}\Bigl(\alpha_i+\frac{V_i}{B_i}\Bigr),$}
\label{eq:comm}
\end{equation}
where $K$ is the number of collectives per step, $\alpha_i$ is the startup and synchronization latency of the $i$-th, $V_i$ is the bytes it moves over the bottleneck link, and $B_i$ is the bandwidth of that link.
The two terms ask for different link properties: bandwidth shrinks $V_i/B_i$, but a decode collective moves only kilobytes, so its time is dominated by $\alpha_i$, and $K$ is large (130 to 484 per step for the models in \S\ref{sec:evaluation}).
Fast links must therefore be low-latency as well as wide, and that latency is what \textbf{Finding 4} measures across four pool sizes, from one chip to eight, on one machine.

\subsection{Economical Resources on Galaxy}
\label{sec:design:galaxy}
This subsection presents Galaxy, the pooled lower-end system, and the resources and links of its pools.

\vspace{0.2em}\noindent \textbf{Resources per chip.}
The Galaxy starts from cheap resources: each of its 32 Blackhole chips carries 32 GB of GDDR6 at 512 GB/s, 210 MB of SRAM, and 664 peak Block FP8 TFLOPS~\cite{tenstorrent2026galaxy,tenstorrent2026blackhole}.
At the \$5000 the machine's list price allocates to a chip, this is 6.4 GB, 102 GB/s, and 133 TFLOPS per \$1000, about twice the memory and bandwidth per dollar of the GPU server (\textbf{Finding 3}).
The links are on the same die: each chip has Ethernet cores that drive its own 400 GbE ports, so a chip reaches its neighbours over its own NoC rather than through a host or a PCIe switch; the measured cluster descriptor shows 8 cabled ports on 20 chips and 10 on the rest.

\begin{figure}[t]
\centering\figinclude{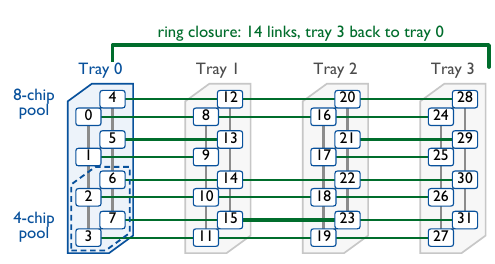}{\columnwidth}
\caption{The communication topology of one Galaxy, from the measured cluster descriptor. Each tray is a $2\times4$ mesh of chips with two links between neighbours (grey, 80 links); the four trays form a ring in which each chip is linked to the same position in the next tray (green, 60 links), and the bracket marks the 14 links that close the ring from tray 3 back to tray 0. An eight-chip pool is one tray and a four-chip pool is a $2\times2$ square inside it.}
\label{fig:galaxy_mesh}\end{figure}

\begin{figure}[t]
\centering\figinclude{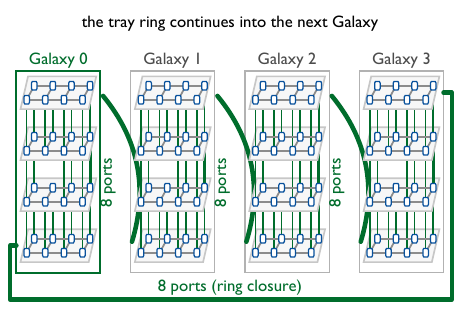}{\columnwidth}
\caption{A four-Galaxy supercluster. The tray ring of each system continues through its edge ports into the next, so the 128 chips form one ring of 16 trays; each green bundle is eight same-position links through 800 GbE QSFP-DD ports.}
\label{fig:galaxy_cluster}\end{figure}

\vspace{0.2em}\noindent \textbf{Pools of chips.}
A pool is the group of chips that serves one model replica, and the Galaxy's wiring lets a pool be any rectangle of its grid (Figure~\ref{fig:galaxy_mesh}).
The 32 chips sit on four trays of eight; within a tray the chips form a $2\times4$ mesh with two links between every neighbouring pair, 80 links in all, and the four trays form a ring in which each chip is linked to the chip at the same position in the next tray, 60 links more, so the whole machine is a $4\times8$ grid with wrap-around along the tray axis.
Galaxy places a pool on any rectangle of this grid: a four-chip pool on a $2\times2$ square, an eight-chip pool on one tray, a 16-chip pool on two trays, and a 32-chip pool on the whole machine; it runs each pool's collectives on a logical ring over its chips and places several pools side by side on disjoint chips. The inter-tray links carry the same 400 GbE per direction, so the ring construction extends across trays and machines unchanged.

\vspace{0.2em}\noindent \textbf{Beyond one machine.}
The same wiring continues across machines, which is what makes the Galaxy a template for the lower-end clusters of \textbf{Finding 2} rather than a single box.
The Ethernet ports at the edge of the tray ring are exposed on the chassis, up to 56 QSFP-DD ports at 800 GbE, and are cabled to the matching ports of the next system, so the tray ring of one Galaxy continues into the next and a four-system supercluster is one ring of 16 trays with no switch on the path (Figure~\ref{fig:galaxy_cluster})~\cite{tenstorrent2026galaxy}.
The vendor offers this configuration from four to more than 36 systems, and Equation~\ref{eq:comm} applies to it unchanged: each added hop adds one link latency to $\alpha_i$ and nothing to the bytes each rank moves, so the communication time grows by one $\alpha$ per machine on the ring and by nothing in its bandwidth term.

\subsection{Fast Links on Galaxy}
\label{sec:design:comm}
This subsection demonstrates the communication system of Galaxy: the links and routers that keep the communication overhead small on every hop of the ring.

\begin{figure}[t]
\centering\figinclude{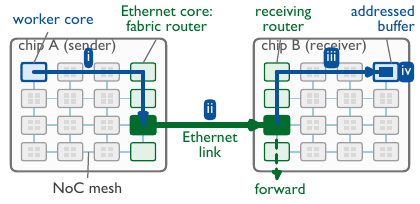}{0.85\columnwidth}
\caption{The path of one transfer between two Galaxy chips. A worker core writes over the on-chip NoC to the fabric router on an Ethernet core \idx{i}, the router sends over the link \idx{ii}, the receiving router delivers to its local NoC or forwards toward the destination \idx{iii}, and the NoC writes the addressed buffer \idx{iv}.}
\label{fig:fabric_path}\end{figure}

\vspace{0.2em}\noindent \textbf{Low startup cost.}
The communication time of Equation~\ref{eq:comm} is spent mostly on the startup term $\alpha_i$, so the links must keep that term small, and the Galaxy does so by keeping the host out of the data path.
Figure~\ref{fig:fabric_path} shows the path of one transfer: a worker core writes a tensor over the on-chip NoC to a fabric router that runs on an Ethernet core; the router sends it over the link, the receiving router forwards it toward its destination or delivers it to the local NoC, and the NoC writes it into the addressed buffer.
The link runs in TT-link mode, in which hardware adds a 16-byte header, places data into receiver buffers, and retransmits with Go-Back-N~\cite{tenstorrent2026fabric}; the host configures the transfer but never moves the bytes.
A 64 KiB all-gather therefore takes 15.0~$\mu$s on a ring of four chips and 21.0~$\mu$s on a ring of eight, against 119 and 257~$\mu$s for the same collectives over NCCL TCP sockets on the GPU server, the reference for the weak links of \S\ref{sec:characterization} and \S\ref{sec:rq2}.

\begin{figure}[t]
\centering\figinclude{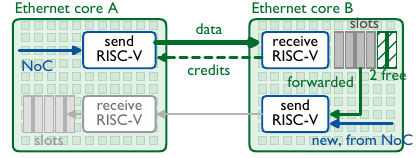}{0.85\columnwidth}
\caption{Inside the Ethernet cores at the two ends of one link. Each core runs a send and a receive RISC-V processor; data crosses the link only against credits returned by the receiver, and a chip injects new traffic only while two of its receiver buffer slots stay free for forwarded traffic, which keeps the ring free of deadlock.}
\label{fig:eth_core}\end{figure}

\vspace{0.2em}\noindent \textbf{Pipelined links.}
The bandwidth term $V_i/B_i$ stays small when transfers overlap on the ring, which the fabric arranges in hardware and firmware.
The installed Blackhole firmware runs separate send and receive RISC-V processors on each Ethernet core, credit-based flow control protects the receiver buffers, and ring deadlock avoidance reserves two free slots for forwarded traffic before a chip injects new traffic (Figure~\ref{fig:eth_core}).
Measured on rings of four and eight chips, a 64 MiB all-gather reaches 85 to 86 GB/s of bus bandwidth, where bus bandwidth is $(n-1)V/T$ for $n$ ranks contributing $V$ bytes each in time $T$, against 5.5 and 4.3 GB/s over the TCP sockets; each timing is the mean of 20 trace replays after five warm-ups with one link per hop.

\vspace{0.2em}\noindent \textbf{Achieved bandwidth.}
What the model sees is the achieved collective bandwidth, which sits below the port rate.
The collective helper chooses how many links a collective uses, and specialized kernels set their own parameters~\cite{tenstorrent2026ccl}, so a collective rarely drives every port: 86 GB/s per chip is about a fifth of the 400 to 500 GB/s its cabled ports could carry.
The transfer term $V_i/B_i$ of Equation~\ref{eq:comm} is set by this software rather than by the ports' full rate, and the headroom above it is a route to a smaller term that needs no new hardware, only firmware and kernel work.

\subsection{Adapting Model Inference to the Links}
\label{sec:design:pool}
This subsection demonstrates how Galaxy adapts model inference to its fast links.
A pool runs one model replica with standard tensor parallelism~\cite{shoeybi2019megatron}, two ring collectives per layer over the links of \S\ref{sec:design:galaxy}, and the adaptation lies in how those collectives are overlapped with compute on the ring and issued without the host, so that the links of \S\ref{sec:design:comm} are the only thing a collective waits for.

\begin{figure}[t]
\centering\figinclude{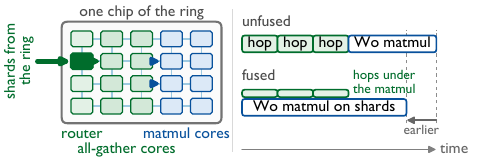}{\linewidth}
\caption{The fused all-gather and matmul on one chip of the ring. The fabric router on the Ethernet core delivers the attention-head shards from the ring neighbour, one set of Tensix cores places them, and another set runs the output projection on each shard as it lands, so the hops of the collective run under the matmul instead of before it, on every chip of the ring at once.}
\label{fig:agmm_overlap}\end{figure}

\vspace{0.2em}\noindent \textbf{Overlapping the all-gather with compute.}
Figure~\ref{fig:fabric_path} separates the two kinds of core a transfer touches: the fabric router on the Ethernet core moves the data, and the Tensix cores are free while it is in flight.
Galaxy's decode uses that separation to hide a collective behind the matmul that needs it.
The all-gather of the attention heads runs on one set of Tensix cores and the output projection on another, and the projection consumes each head shard as the router delivers it from the ring, so the $n-1$ hops of the collective run under the matmul instead of before it (Figure~\ref{fig:agmm_overlap}).
This adaptation takes one of the two collectives per layer off the critical path: its $\alpha_i$ and $V_i/B_i$ in Equation~\ref{eq:comm} are paid under compute, and the step sees one collective per layer on its critical path instead of two, on every chip of the pool.

\begin{figure}[t]
\centering\figinclude{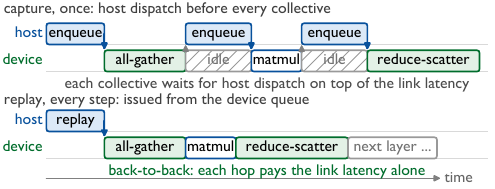}{\linewidth}
\caption{Capturing and replaying one decode step. The capture pass runs once at start-up, dispatched eagerly from Python: the device waits for the host to prepare and enqueue each operation, so the ring idles for a host dispatch before every collective. Every later decode step, for every request, replays the captured operations from the device queue back-to-back, so each hop pays the link latency alone, with no host in the loop between one collective and the next one on the ring, step after step.}
\label{fig:trace_replay}\end{figure}

\vspace{0.2em}\noindent \textbf{Host-free decode steps.}
The links of \S\ref{sec:design:comm} make a 64 KiB all-gather cost 15 to 21~$\mu$s on a ring, but a decode step issues hundreds of collectives, and if the host enqueued each of them from Python the dispatch would be paid on every one, on top of the link latency.
Galaxy therefore runs decoding through Metal Trace~\cite{tenstorrent2025trace} in two sub-steps, \idx{i} capture and \idx{ii} replay.
For the capture, Galaxy runs one decode step eagerly at start-up while Metal Trace records every operation and collective it issues.
For the replay, every later decode step issues that recording from the device's command queue (Figure~\ref{fig:trace_replay}): the collectives run back-to-back, and none waits for the host to dispatch it.
This adaptation works on the startup term $\alpha_i$ of Equation~\ref{eq:comm}: each hop pays the link latency of \S\ref{sec:design:comm} and no host time, so a decode step is as short as the fabric and the kernels allow, whatever the host is doing in the meantime.

\section{Implementation}
\label{sec:implementation}
\vspace{0.2em}\noindent \textbf{Serving stack.}
Galaxy serves through TT inference server 0.25.0~\cite{tenstorrent2026inference} on TT-Metal revision \texttt{978f93b56e36}~\cite{tenstorrent2026ttmetal} and firmware 19.11.0.
The server's Llama and Qwen model code shards each layer across the chips of a pool~\cite{shoeybi2019megatron}, keeps the KV cache in paged device memory~\cite{kwon2023pagedattention} with the heads that own it, and runs decode as a captured Metal Trace~\cite{tenstorrent2025trace} that every later step replays, with the next token selected on the device.
The fabric is brought up in \texttt{FABRIC\_1D\_RING} mode~\cite{tenstorrent2026fabric}, one logical ring per pool over the links of \S\ref{sec:design:galaxy}, with every hop forwarded by the Ethernet-core routers of \S\ref{sec:design:comm} rather than by a switch.

\vspace{0.2em}\noindent \textbf{Communication system.}
Every collective runs on the Ethernet cores of \S\ref{sec:design:comm}, whose router firmware forwards packets around the ring under credit-based flow control and the two-slot injection rule~\cite{tenstorrent2026fabric}.
The model issues them through TT-NN's asynchronous kernels~\cite{tenstorrent2026ccl}, the ring all-gather, the minimal reduce-scatter that implements every all-reduce, and the fused all-gather-matmul of \S\ref{sec:design:pool}, each tuned per model and pool size in links per hop, worker cores per link, and chunks between synchronizations.
Payloads cross the ring in block floating point and land in the width-sharded layout the next matmul reads, and banks of hardware semaphores, double-buffered per ring axis, let back-to-back collectives proceed without waiting on one another~\cite{tenstorrent2026ccl}.

\section{Evaluation}
\label{sec:evaluation}

We design the evaluation to answer two research questions:
\begin{itemize}[leftmargin=*]
\item \textbf{\underline{RQ1:}} \emph{How does the Galaxy compare with the GPU server in tokens per dollar and tokens per joule?}
\item \textbf{\underline{RQ2:}} \emph{How much of that efficiency do slower links take away at the same active concurrency?}
\end{itemize}

\subsection{Experimental Setup}
\label{sec:setup}

\noindent \textbf{Hardware environment.}
The lower-end cluster is one Tenstorrent Blackhole Galaxy, 32 chips of 32\,GB each joined by the Ethernet links of \S\ref{sec:design:galaxy}, with the serving stack of \S\ref{sec:implementation}~\cite{tenstorrent2026galaxy}.
The high-end baseline is one server with eight NVIDIA RTX PRO 6000 Blackwell Server Edition GPUs (96\,GB each) connected over PCIe~5.0~$\times$16 links inside a single chassis~\cite{nvidia2025rtxpro6000}.

\vspace{0.2em}\noindent \textbf{Models and workloads.}
We evaluate Llama-3.3-70B-Instruct, Qwen3-32B, and Llama-3.1-8B-Instruct in BF16 and 8-bit configurations~\cite{grattafiori2024llama3,yang2025qwen3}; the 8-bit runs use Galaxy BFP8 and GPU FP8 weights and KV cache, formats that differ numerically.
We test workloads whose context length varies from 8K to 256K tokens, so that each comparison spans short sessions and long resident ones; K denotes 1024 tokens.

\vspace{0.2em}\noindent \textbf{Baseline systems.}
We compare three systems, and the same version of vLLM~\cite{kwon2023pagedattention} serves on top of each.
\idx{i}~GPU server: the RTX PRO 6000 Blackwell server described above.
\idx{ii}~Galaxy with our communication system: the chips pooled over their native links by the communication system of \S\ref{sec:design}.
\idx{iii}~Galaxy with TCP-class links: the same chips and batches with every collective raised to the bandwidth of NCCL over TCP sockets, modelling the slow links between machines in a conventional cluster (\S\ref{sec:rq2}).

\vspace{0.2em}\noindent \textbf{Metrics.}
We report the two cluster-efficiency metrics of Equation~\ref{eq:cost}, tokens per dollar and tokens per joule.
Tokens per dollar uses the selling price of each system, the list price of the Galaxy and the market price of the RTX PRO 6000 Blackwell server~\cite{tenstorrent2026galaxy,punch2026rtxpro6000server,bytehaus2026aiservers}.
We record power through the core-rail telemetry of the Tenstorrent system management interface and the board telemetry of the NVIDIA management library, sampled every one to two seconds throughout every timed round~\cite{tenstorrent2026ttsmi,nvidia2025nvml}.
The two telemetries have different boundaries in the sense of Equation~\ref{eq:cost}: the core-rail reading covers the supply of each chip's Tensix cores, network-on-chip, and Ethernet cores, while the board reading covers the whole GPU board.
The power outside the core rail varies with the workload, so Table~\ref{tab:sensitivity} in \S\ref{sec:rq1} reports the sensitivity of tokens per dollar to Galaxy power above the core-rail reading.

\subsection{End-to-End Efficiency Evaluation (RQ1)}
\label{sec:rq1}

\begin{figure*}[!t]
\centering\figinclude{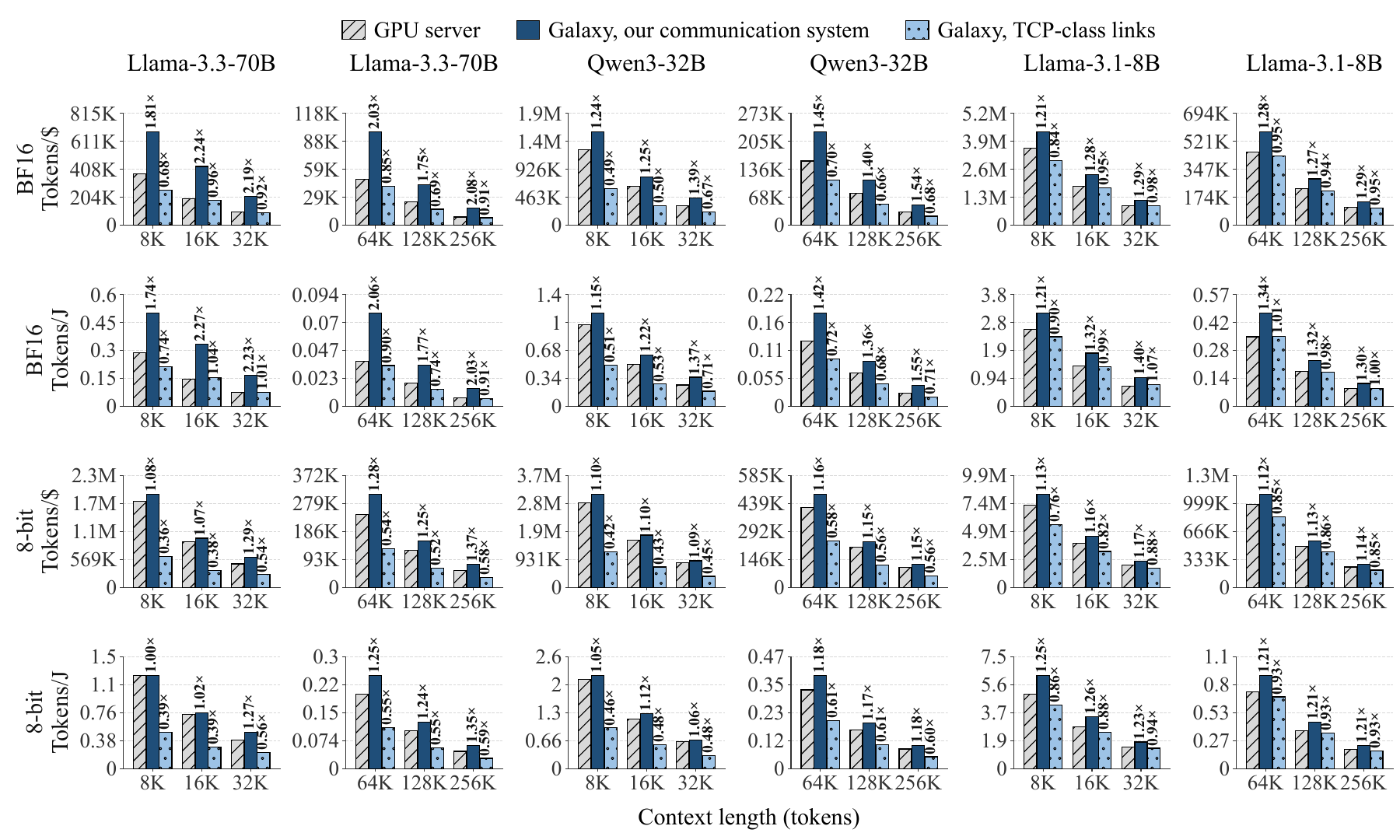}{\textwidth}
\caption{Tokens per dollar and tokens per joule of the GPU server, of the Galaxy with our communication system, and of the Galaxy over TCP-class links, for three models in BF16 and 8-bit over 8K-256K contexts. Over TCP-class links every collective is raised to the bandwidth and latency of NCCL over TCP sockets, emulated on the native fabric with the same chips and batches.}
\label{fig:ablation}
\end{figure*}

\noindent \textbf{Tokens per dollar.}
The Galaxy delivers more tokens per dollar than the GPU server in all 36 configurations (Figure~\ref{fig:ablation}), by 1.07-2.24$\times$.
The gain is largest for Llama-3.3-70B in BF16, 1.75-2.24$\times$, where the memory of the GPU server limits concurrency: at 16K the Galaxy serves 92 active sequences against 32 on the GPU server and reaches 429K tokens per dollar against 192K.
Qwen3-32B BF16 gains 1.24-1.54$\times$ and Llama-3.1-8B BF16 1.21-1.29$\times$.
Within a series the gain grows with context for the two larger models, from 1.08$\times$ at 8K to 1.37$\times$ at 256K for 70B 8-bit and from 1.24$\times$ to 1.54$\times$ for 32B BF16, as longer contexts leave the GPU server fewer sequences in memory to batch at each decode step than the Galaxy keeps in flight.
Together, these results show that the pooled chips turn their larger pooled memory, bandwidth, and compute into more tokens per dollar at every model, precision, and context length.

\vspace{0.2em}\noindent \textbf{Tokens per joule.}
The Galaxy also delivers more tokens per joule in all 36 configurations, by 1.00-2.27$\times$, with the same ordering across series: 1.74-2.27$\times$ for 70B BF16, 1.15-1.55$\times$ for 32B BF16, 1.21-1.40$\times$ for 8B BF16, and 1.00-1.35$\times$, 1.05-1.18$\times$, and 1.21-1.26$\times$ for the three 8-bit series.
The one configuration at parity is 70B 8-bit at 8K, where 520 active sequences on the 32 Galaxy chips and 324 on the eight GPUs produce 1.270 and 1.269 tokens per joule.
For instance, 70B BF16 at 16K yields 0.333 tokens per joule on the Galaxy and 0.147 on the GPU server, and 8B 8-bit at 32K yields 1.786 against 1.446.
Together, the two metrics show that the lower-end cluster, pooled over its own links, beats the high-end server on both measures a provisioner pays for, the native half of the claim of \S\ref{sec:intro}.

\begin{figure}[!t]
\centering\figinclude{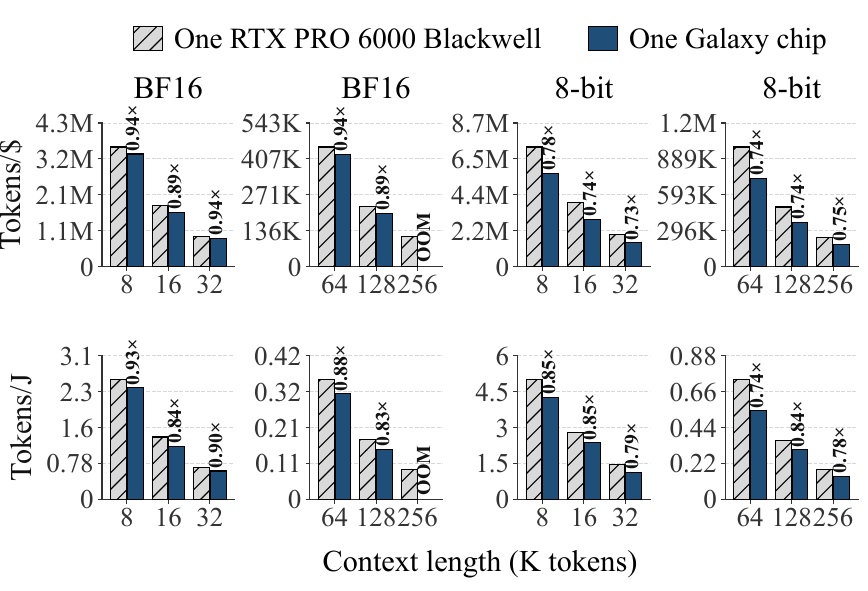}{\columnwidth}
\caption{Tokens per dollar and tokens per joule of one Galaxy chip and one RTX PRO 6000 Blackwell serving Llama-3.1-8B in BF16 and 8-bit over 8K-256K contexts, each at its fastest batch. Tokens per dollar divides each system's price by its device count.}
\label{fig:single}
\end{figure}

\vspace{0.2em}\noindent \textbf{One chip against one card.}
One Galaxy chip stays below one RTX PRO 6000 Blackwell on both metrics in every configuration that fits (Figure~\ref{fig:single}).
The chip serving Llama-3.1-8B holds 0.21-0.25$\times$ the sequences of the card and generates 0.13-0.17$\times$ its tokens per second, and with the price of each system divided by its number of devices it reaches 0.73-0.94$\times$ the card's tokens per dollar and 0.74-0.93$\times$ its tokens per joule in the 11 configurations that fit, and runs out of memory at 256K in BF16.
Together, these results show that the lower-end device loses on device efficiency while the pooled chips of Figure~\ref{fig:ablation} win on cluster efficiency, so the gain comes from pooling over the fast links, not from the chips on their own.

\vspace{0.2em}\noindent \textbf{KV-cache capacity.}
Pooling multiplies the KV-cache capacity of the chips faster than that of the cards and keeps it cheaper per token at every pool size (Table~\ref{tab:kv}).
One Galaxy chip holds 135K tokens of Llama-3.1-8B context in BF16 and 349K in 8-bit, 0.24-0.28$\times$ one RTX PRO 6000 Blackwell, because the weights take 16\,GB of its 32\,GB in BF16 and 8\,GB in 8-bit.
Pooling spreads the weights: a model replica on eight chips holds 1.93M tokens in BF16 and 3.73M in 8-bit, 14.3$\times$ and 10.7$\times$ one chip, while a model replica on eight cards holds 5.57M and 11.2M, 9.7$\times$ and 9.0$\times$ one card.
Per dollar of acquisition price, with the price of each system divided by its device count, the chips hold 27-48 KV tokens in BF16 and 70-93 in 8-bit against 20-25 and 45-50 for the cards.
Together, these results show that the chips hold 1.3-1.9$\times$ more context per dollar than the cards, and that the gap widens with pool size.

\begin{table}[ht]
\centering
\caption{KV-cache capacity of one Llama-3.1-8B replica on one, two, four, and eight Galaxy chips and RTX PRO 6000 Blackwell cards, in tokens and in tokens per dollar of acquisition price, with the price of each system divided by its device count; the better of chips and cards at each device count is set in bold.}
\label{tab:kv}
\setlength{\tabcolsep}{3pt}
\renewcommand{\arraystretch}{0.95}
\resizebox{\columnwidth}{!}{%
\small\begin{tabular}{l | r | r | r | r | r | r | r | r}
\hline
 & \multicolumn{4}{c|}{BF16} & \multicolumn{4}{c}{8-bit} \\
\hline
 & \multicolumn{2}{c|}{KV tokens} & \multicolumn{2}{c|}{KV tokens/\$} & \multicolumn{2}{c|}{KV tokens} & \multicolumn{2}{c}{KV tokens/\$} \\
\hline
Devices & chips & cards & chips & cards & chips & cards & chips & cards \\
\hline
1 & 135K & \textbf{572K} & \textbf{27} & 20 & 349K & \textbf{1.25M} & \textbf{70} & 45 \\
\hline
2 & 383K & \textbf{1.29M} & \textbf{38} & 23 & 832K & \textbf{2.70M} & \textbf{83} & 48 \\
\hline
4 & 899K & \textbf{2.72M} & \textbf{45} & 24 & 1.79M & \textbf{5.55M} & \textbf{90} & 49 \\
\hline
8 & 1.93M & \textbf{5.57M} & \textbf{48} & 25 & 3.73M & \textbf{11.24M} & \textbf{93} & 50 \\
\hline
\end{tabular}}
\end{table}

\vspace{0.2em}\noindent \textbf{Scaling with pool size.}
Over fast links, every chip added to a replica raises the Galaxy's efficiency per chip as well as per joule (Table~\ref{tab:scaling}).
For one replica of Llama-3.1-8B at 32K context, throughput per chip rises monotonically from 81 to 132 tokens per second in 8-bit and from 47 to 70 in BF16 as the replica grows from one chip to eight, 1.6$\times$ and 1.5$\times$, and tokens per joule from 1.14 to 1.80 and from 0.62 to 1.01.
Pooling the chips' memory, bandwidth, and compute does three things at once: the weights' share of each chip's memory shrinks, so a larger part of the pool goes to KV cache, which long-context agent serving needs for the resident sessions whose prefixes it reuses~\cite{zhu2026tracelab,zheng2024sglang,qin2025mooncake}; the memory I/O and compute available to the replica grow with the chip count; and the memory-bound decode step batches more sequences.
Together, these results show that with fast links the Galaxy keeps gaining from every chip added to a replica, in tokens per second per chip and in tokens per joule.

\begin{table}[ht]
\centering
\caption{Throughput per chip and tokens per joule of one Llama-3.1-8B replica on one, two, four, and eight Galaxy chips at 32K context; the best value in each column is bold.}
\label{tab:scaling}
\setlength{\tabcolsep}{3pt}
\renewcommand{\arraystretch}{0.95}
\resizebox{\columnwidth}{!}{%
\small\begin{tabular*}{\columnwidth}{@{\extracolsep{\fill}}l | r | r | r | r@{}}
\hline
 & \multicolumn{2}{c|}{BF16} & \multicolumn{2}{c}{8-bit} \\
\hline
Chips & tok/s per chip & tok/J & tok/s per chip & tok/J \\
\hline
1 & 47.4 & 0.62 & 80.6 & 1.14 \\
\hline
2 & 57.0 & 0.83 & 116.5 & 1.65 \\
\hline
4 & 64.6 & 0.97 & 129.4 & 1.78 \\
\hline
8 & \textbf{69.9} & \textbf{1.01} & \textbf{131.9} & \textbf{1.80} \\
\hline
\end{tabular*}}
\end{table}

\vspace{0.2em}\noindent \textbf{Sensitivity to the power boundary.}
Table~\ref{tab:sensitivity} tests how much of the lead in tokens per dollar survives if the Galaxy draws more power than its core-rail telemetry reports (\S\ref{sec:setup}): it multiplies the Galaxy's measured core-rail power by $m$ from 1 to 3 and keeps the GPU board power as measured.
Tripling the core-rail power lowers the average ratio of every series by 10\%: at $m=3$ the BF16 series still average 1.82$\times$, 1.25$\times$, and 1.15$\times$ for 70B, 32B, and 8B, and the three 8-bit series 1.01-1.10$\times$, so all six averages stay above parity.
The drop is small because the tokens per dollar $D$ of Equation~\ref{eq:cost} divide by the purchase price $C$ plus the electricity term $kP$, and at the measured device power $P$ the electricity term is only 5-6\% of $C+kP$, so tripling $P$ raises $C+kP$ by only 10-12\%.
Together, these results show that the lead in tokens per dollar rests on the purchase price of the pooled chips rather than on the boundary of their power telemetry.

\begin{table}[ht]
\centering
\caption{Tokens per dollar of the Galaxy relative to the GPU server, averaged over the contexts of each series of Figure~\ref{fig:ablation}, with the Galaxy's measured core-rail power multiplied by $m$ to stand for the power outside that rail and the GPU board power unchanged.}
\label{tab:sensitivity}
\setlength{\tabcolsep}{3pt}
\renewcommand{\arraystretch}{0.95}
\resizebox{\columnwidth}{!}{%
\small\begin{tabular*}{\columnwidth}{@{\extracolsep{\fill}}l | r | r | r | r | r | r@{}}
\hline
 & \multicolumn{2}{c|}{Llama-3.3-70B} & \multicolumn{2}{c|}{Qwen3-32B} & \multicolumn{2}{c}{Llama-3.1-8B} \\
\hline
$m$ & BF16 & 8-bit & BF16 & 8-bit & BF16 & 8-bit \\
\hline
1 & 2.01$\times$ & 1.22$\times$ & 1.38$\times$ & 1.12$\times$ & 1.27$\times$ & 1.14$\times$ \\
\hline
1.5 & 1.96$\times$ & 1.19$\times$ & 1.34$\times$ & 1.09$\times$ & 1.24$\times$ & 1.11$\times$ \\
\hline
2 & 1.91$\times$ & 1.16$\times$ & 1.31$\times$ & 1.07$\times$ & 1.21$\times$ & 1.08$\times$ \\
\hline
2.5 & 1.87$\times$ & 1.13$\times$ & 1.28$\times$ & 1.04$\times$ & 1.18$\times$ & 1.06$\times$ \\
\hline
3 & 1.82$\times$ & 1.10$\times$ & 1.25$\times$ & 1.01$\times$ & 1.15$\times$ & 1.03$\times$ \\
\hline
\end{tabular*}}
\end{table}
\subsection{Communication-Link Ablation (RQ2)}
\label{sec:rq2}

\noindent \textbf{Throughput loss.}
Connecting the same chips over TCP-class links cuts their throughput by 23.5-67.4\% across the 36 Galaxy configurations, 47.3\% on average, with the hardware capacity, the batches, and the acquisition cost unchanged.
The 70B and 32B models lose 50.4-67.4\% and the 8B model 23.5-32.6\%; the larger models issue 257-484 collectives per decode step against 130-131 for 8B, so the added startup latency is paid more often for every token generated.
Together, these results show that the link cost alone, with nothing else changed, takes away between a quarter and two thirds of the throughput that the pooled chips deliver.

\vspace{0.2em}\noindent \textbf{Efficiency.}
The loss erases the native advantage: over TCP-class links the Galaxy falls below the GPU server in tokens per dollar in every one of the 36 configurations and in tokens per joule in 31 of the 36 (Figure~\ref{fig:ablation}).
The ratios are 0.36-0.98$\times$ and 0.39-1.07$\times$; the five points that stay at or above parity are 8B and 70B BF16 configurations where the loss is smallest and the pool draws less power while it waits on the links.
For instance, 70B BF16 at 16K drops from 429K to 184K tokens per dollar, 0.96$\times$ the GPU server, and 70B 8-bit at 8K from 1.90M to 0.62M, 0.36$\times$.
Together, these results show that, since nothing but the link cost changes between the two Galaxy bars of Figure~\ref{fig:ablation}, the fast links are what let the lower-end chips keep their per-dollar and per-joule advantage once pooled, as \S\ref{sec:intro} claims.
\section{Conclusion}
This paper asks how an LLM serving cluster should be built when the goal is cluster efficiency, the tokens it serves per dollar and per joule, rather than device efficiency.
Across three models, two precisions, and contexts from 8K to 256K, the Galaxy delivers more tokens per dollar and more tokens per joule than an RTX PRO 6000 Blackwell server in all 36 configurations, by 1.07-2.24$\times$ and 1.00-2.27$\times$, while a single Galaxy chip stays below a single GPU on both metrics.
Connecting the same chips over TCP-class links cuts their throughput by 23.5-67.4\% and erases all 36 tokens-per-dollar wins and 31 of the 36 tokens-per-joule wins.
The advantage of a lower-end cluster comes from pooling over fast links: start from economical accelerators, connect them with fast links, and judge the cluster by what it serves per dollar and per joule rather than by any one device.

\bibliography{references}
\bibliographystyle{mlsys2025}

\end{document}